# Directing Space: Rehearsing Architecture as Performer with Explainable AI

Pavlos Panagiotidis*[1][0009-0000-8229-7239] Jocelyn Spence[1][0000-0002-4322-1394] Nils Jaeger[2][0000-0002-4686-2568] Paul Tennent[1][0000-0001-6391-0835]

[1] School of Computer Science, University of Nottingham, Nottingham, United Kingdom
[2] Department of Architecture and Built Environment, Faculty of Engineering, University of Nottingham, Nottingham, United Kingdom

*Corresponding author: `pavlos.panagiotidis@nottingham.ac.uk`

`{jocelyn.spence4, nils.jaeger, paul.tennent}@nottingham.ac.uk`

**Abstract.** As AI systems increasingly become embedded in interactive and immersive artistic environments, artists and technologists are discovering new opportunities to engage with their interpretive and autonomous capacities as creative collaborators in live performance. The focus of this work-in-progress is on outlining conceptual and technical foundations under which performance-makers and interactive architecture can collaborate within rehearsal settings. It introduces a rehearsal-oriented prototype system for shaping and testing AI-mediated environments within creative practice. This approach treats interactive architecture as a performative agent that senses spatial behaviour and speech, interprets these signals through a large language model, and generates real-time environmental adaptations. Designed for deployment in physical performance spaces, the system employs virtual blueprints to support iterative experimentation and creative dialogue between artists and AI agents, using reasoning traces to inform architectural interaction design grounded in dramaturgical principles.

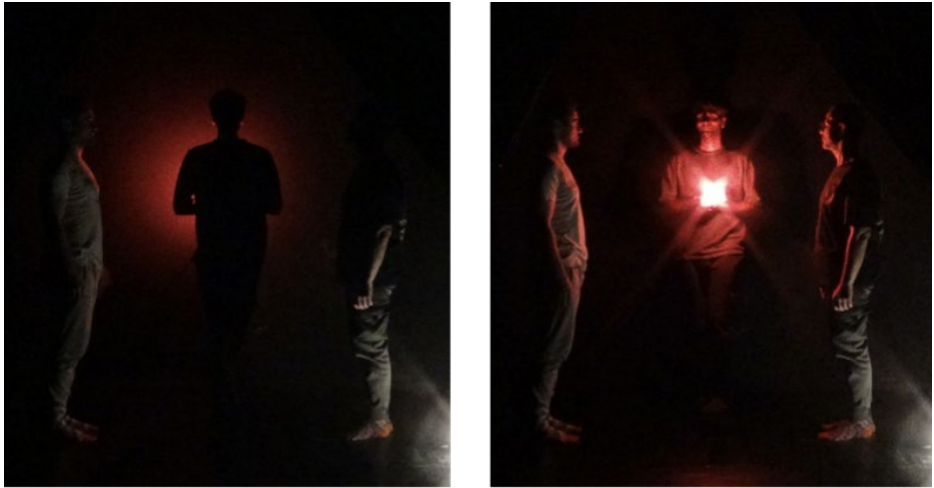

**Fig. 1.** Responsive light adaptation. A prototype demonstrating how the system triggers light adaptation based on human spatial relationships during rehearsal.

## 1      Introduction

In the novel *The Thousand Dreams of Stellavista* [1], houses absorb the emotional residues of their occupants and respond in kind; the story offers both a critique of technological progress and a provocation to imagine architecture as a reactive character shaped by human behaviour. This idea resonates with contemporary immersive performance, where architecture functions as dramaturgical material: environments frame and guide experience, surrounding audiences who often move freely through space [2, 3], echoing how architecture can be understood less by what it is than by what it does [4]. Building on this view, we examine how architecture can perform autonomously through AI mediated decision making. Beyond its creative appeal, this approach leverages the capacity of AI structures to handle complex, large-scale input and output flows, coordinating data from multiple rooms, managing numerous actuators in parallel, and analysing long term behavioural patterns. While such systems may lack the subtlety of human interpretation, they afford levels of scalability and continuity unattainable through human operation alone.

We introduce an AI-driven virtual blueprint that guides spatial behaviour through dialogic, dramaturgical interaction, shifting responsive environments from reactive automation toward interpretable, co-creative architectural agency. The blueprint integrates an LLM-based architectural agent into a virtual environment, allowing system behaviour to be shaped through natural language rather than hard-coded rules. This setup allows behavioural tendencies to be established and refined before they are transferred into the physical environment. The current prototype realises the system's foundational elements, using a large language model to interpret dramaturgical input and generate spatial adaptations in a virtual built environment. Additionally, early physical trials link sensors and actuators to their virtual counterparts. As the project progresses, we will examine how artists engage with such systems, how dramaturgical meaning emerges in dialogue with LLM agents, and how reasoning traces can support creative collaboration. Although developed for performance making, the approach also points toward broader applications in socially responsive environments where spaces participate in interaction and co-design processes with their users.

## 2    Background and Conceptual Foundations

This section outlines the conceptual and methodological foundations of the work. It draws on theatre-making, mixed-reality performance practices and research in virtual scenography. In this context, performance refers to immersive and interactive events where the actions of performers, audiences, and the built environment are witnessed and responded to. The work does not focus on traditional on-stage performance but rather on expanded forms of experiences with theatrical qualities.

Research at the intersection of performance and technology has shown how computation can act as a dramaturgical force, framing interaction as performative and technologies as expressive agents [5–8]. Artistic companies have similarly explored mixed-reality environments where spatial design, adaptive media, and networked technologies are integral to audience experience [9–11].

Building on this trajectory, our framework brings dramaturgical logics into rehearsal within mixed reality systems where theatre-makers collaborate in real time with AI agents that influence the built environment. It extends a previous study that developed a no-code, rule-based authoring tool for responsive environments in a devising process [12]. Devising is a collaborative, improvisation-based method of theatre-making where performances emerge through exploration rather than a predetermined script. The tool was designed to make experimentation with responsive spaces faster and more intuitive, reducing reliance on technologists and keeping creative iteration within rehearsal. Performers linked gestures, positions, and vocal expressions sensed through computer vision and audio analysis to architectural adaptations, triggering scenographic responses such as light changes based on predefined mappings (Figure 1). This enabled theatre devisers to explore how embodied relations could shape spatial behaviour, though responsiveness remained limited to explicitly defined rules. The current system aims to address this limitation by allowing theatre-makers to express intentions in natural language, with the LLM improvising responses in the form of environmental adaptations.

Yet enabling such collaboration in practice introduces several methodological and practical challenges, as performance-making is often limited by the scale and cost of building scenographic environments solely to test interaction logic. Virtual environments offer a way to explore possibilities without premature physical construction. Prior work in 3D, VR, and digital twin technologies shows how virtual models can function as prototypes [13, 14] and how collaborative environments can function as shared rehearsal spaces [15–17]. In our approach, simulation enables experimentation with interactions guided by the AI agent itself. While Wizard-of-Oz methods [18] can only approximate system behaviour, a virtual model with an embedded AI agent allows the interaction logic to be shaped and rehearsed before deployment. Once the creative team settles the desired behaviour in the virtual prototype, this logic can be transferred to the physical setup by linking real sensors and actuators to their virtual counterparts, allowing the interaction to continue in the physical environment.

For AI agents to take part meaningfully in rehearsal, they need to perceive events, simulated or real, as having dramaturgical significance. They should be able to recognise them in a way that at least comes close to how human collaborators experience them in live rehearsal. The problem is that many performance practices involve subtle, layered behaviours that are not easy for machines to read. For example, frameworks such as *Laban Movement Analysis* can be used to describe movement qualities [19], but the very nature of these qualities makes them hard to translate into computational terms [20]. For this reason, we turn to two performance-making methods that include actions and events which, although cannot be fully reduced to data, they can make observation easier for the kinds of AI systems we have today. We refer to examples such as spatial relationships, topographies, and gestures from *Viewpoints* [21], as well as zones of attention, voids, and gravitational points from immersive theatre-making [3].

Furthermore, we draw on the foundational rehearsal logic developed by Stanislavski, which includes backstories, objectives, and given circumstances to support the dramaturgical framing of interactions [22, 23]. Although it is perhaps one of the most subtle and complex dramaturgical frameworks, we see potential in how it can be expressed through directing instructions, discussions, and rehearsal notes. This means that it may help form a basis for describing dramaturgical frames to the LLM-based system through language, although how far this can go remains to be seen in practice.

Our system operationalises these theatrical logics through three components:

1. a sensing layer that perceives behaviour, using overhead computer vision for position tracking with YOLOv8 [24] and speech analysis on recorded audio using Vosk [25]
2. a large language model accessed through OpenAI's 4o Model API [26] that interprets dramaturgical intent from open ended directions
3. a Unity [27] based virtual environment that functions as a rehearsal space for testing and refining interactions before physical deployment, currently connected to Philips Hue smart lighting for environmental actuation [28].

When the system makes choices—currently limited to changing lighting configurations with potential extensions to sound cues or moving objects—an explainability layer exposes its reasoning through a textual trace. This trace allows directors and designers to review decisions and adjust dramaturgical inputs in an iterative, "ping-pong" rehearsal between human and system. It addresses a common problem in interactive AI, where opaque system behaviour hinders collaboration [29], by keeping reasoning visible. It remains to be seen to what extent the reasoning trace supports the creative process, since it functions as a post hoc rationalisation of the system's decisions rather than a direct account of its reasoning. A later stage of this study will focus on how such explanations, even if partial or approximate, can assist creative practitioners in shaping and directing the system's behaviour. Taken together, these components form the groundwork for an LLM-mediated interactive architectural system for immersive performance-making. The following section outlines the system's structure and operation.

## 3  Interaction Generation Flow

### 3.1  Staging a Collaborative System

The system is presented here as an evolving prototype developed in *Unity* [27]. We outline both its current capabilities and its intended future development. It is capable of perceiving real or simulated data and triggering responses in the virtual blueprint, and eventually in the physical space. Its capacity to operate in a rehearsal simulation mode, where virtual agents and synthetic speech emulate performer and audience behaviour (Figure 2, top left), is designed to allow creatives to explore and refine how the AI responds to dramaturgical framing. It enables them to test different scenarios and observe its reactions without the constant need for live participants or physical setups. Position data are also aggregated into a dynamic heat grid that visualises patterns of movement and attention, allowing both system and users to detect zones of concentrated activity or "hotspots" (Figure 2, top right).

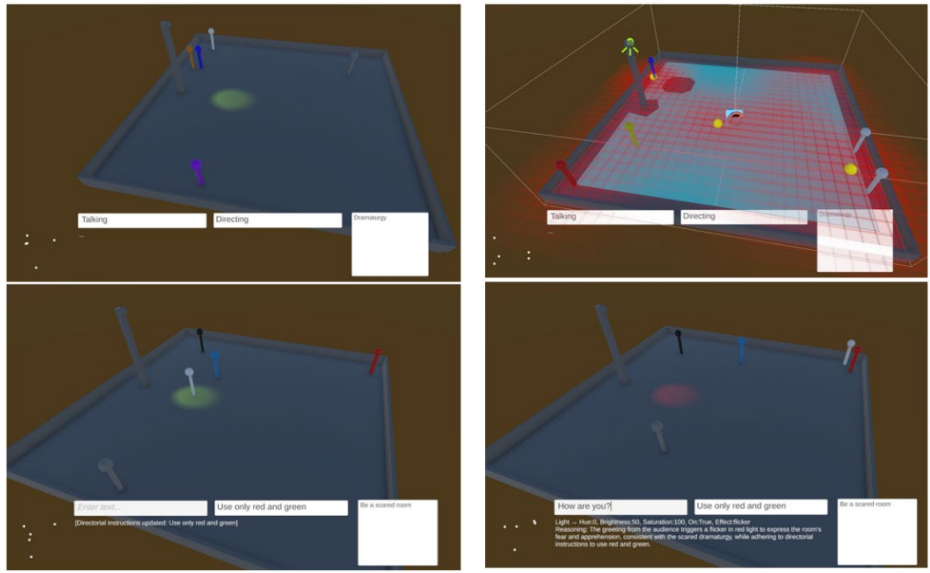

**Fig. 2.** System responses and visualization modes. Top left: virtual lighting cue where a green light activates when a virtual spectator approaches the pillar. Top right: audience heatmap with red indicating high activity, blue low, and yellow marking gravitational points. Bottom left: dramaturgical or directorial prompts where the system is instructed to "be a scared room" and "use only red and green." Bottom right: LLM reasoning trace showing the system's response to "How are you?" through a red light and the explanation "The greeting triggers red light to express fear, consistent with the scared dramaturgy and red–green instruction."

Beyond the sensing and the actuation layers, the system includes a dramaturgical layer through which the designer or director defines a creative frame to guide its behaviour. While explicit behavioural rules can still be specified, this approach feels largely exhausted, conceptually if not technically. Instead of just prescribing actions, this layer can be used to offer a short narrative or contextual prompt that gives the environment a sense of prior experience, motivation, or emotional tone—effectively directing the system's "inner world" and assigning the AI a dramaturgical standpoint from which to generate interaction (Figure 2, bottom left). When the system makes a decision, for example changing a light because a participant said something, it explains its reasoning (Figure 2, bottom right) to help the practitioner refine its behaviour through further prompting.

These layers (real and simulated inputs, decision engine, and actuators) are already functional within the current prototype. In its full configuration, the AI will integrate the dialogic interaction described in the following sections to deepen the dramaturgy guiding the system's behaviour in live rehearsal.

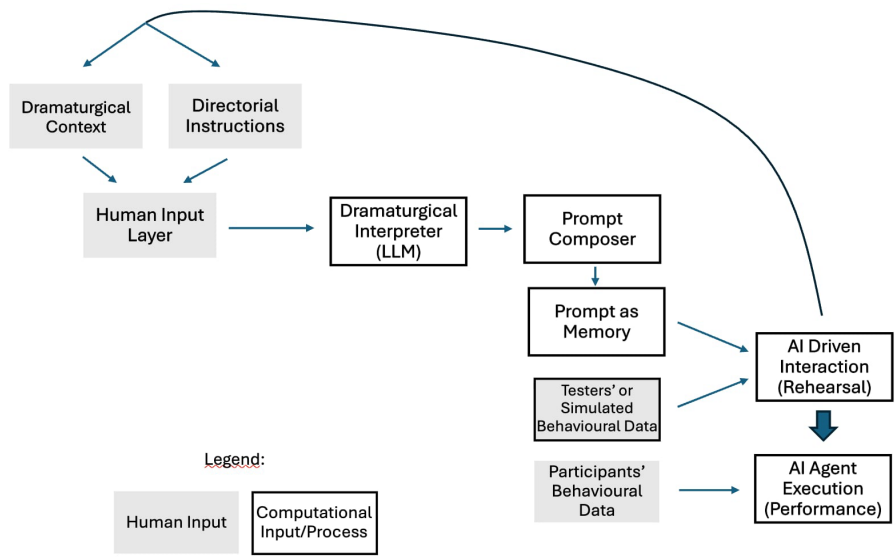

**Fig. 3.** Interaction generation pipeline. The system synthesises dramaturgical and directorial inputs, interpretive analysis, and accumulated rehearsal memory into a single consolidated prompt defining the AI agent's performative behaviour.

### 3.2   Dramaturgical Interpreter

Building on the dramaturgical layer, the system is being extended with an interpreter that translates open-ended dramaturgical descriptions into more concrete parameters. For example, a director might prompt: "You are the house from *The Thousand Dreams of Stellavista* [1]. You have witnessed turbulent relationships, absorbed traces of jealousy and loss, and now respond cautiously to emotional tension in the room." Such framing does not prescribe actions but situates the system within a narrative perspective.

The interpreter analyses the text to extract dramaturgically relevant features—objectives ("protect itself from emotional harm"), affective tones ("cautious, melancholic"), or environmental metaphors ("the space tightens when tension rises"). It may also engage the director in brief clarification, for instance asking whether the room's caution should be expressed as a withdrawal of light or as tightening illumination. It will also be interesting to see how different LLM models handle this kind of dialogue, and how much guidance or fine-tuning they might need to engage meaningfully with dramaturgical direction. Each model may approach the dialogue differently, deciding what to focus on and what to leave out, and it will be important to observe how well they can use what they find to inform the actuation. Internally, this information is stored in a lightweight data schema that defines relationships between intention, affect, and spatial behaviour, as shown below (for illustrative purposes):

- "intention": "respond cautiously to emotional tension"
- "affect": " subdued calm",
- "metaphor": "tightening space",
- "primary_modality": "light",
- "reaction_pattern": "gradual dimming under high energy"

### 3.3    Directorial Layer

Sitting above the dramaturgical interpreter, the directorial layer provides more precise control over the system's behaviour. Unlike the dramaturgical layer, which invites interpretation, this mode aims to treat each instruction as a concrete command to be executed. The director may, for example, specify: "When someone enters a room, fade all side lights to 30 percent intensity," or "Trigger the fan when two participants stand within two metres of each other." In this mode, the system does not infer broader meaning but focuses solely on the accurate execution of the given cues.

A potentially useful application of this layer may emerge during live performance. When a director, orchestrator, or even a performer experiences an unexpected situation that requires the environment to adjust in a specific way, this layer can be used in real time to issue clear, language-based instructions to the system. For instance, if the system turns off the lights in a certain area, the director may command, "Turn the lights back on in Room B," prompting an immediate response. It is important to note that, in such cases, the purpose is not to override the system (or replace any dedicated safety overrides that may be in place) but to maintain the behaviour within the logic of the spatial performer system.

### 3.4    Dramaturgical Memory

In rehearsal, the two layers operate together. During, or after each run, the director or designer reviews the system's reasoning traces and provides new instructions, much like in a traditional rehearsal, indicating what worked and what needs adjustment. These notes are incorporated into the next iteration of the loop, guiding the AI to refine its dramaturgical policies. Through successive cycles, the human and system co-evolve a shared vocabulary of interaction, and the environment gradually develops a dramaturgical memory (Figure 3).

This memory comprises accumulated instructions, distilled notes, and behavioural traces from rehearsal runs, capturing both the dialogue between director and system and the patterns emerging through interaction. Rather than storing information indiscriminately, the system filters and organises data according to dramaturgical relevance and the director's guidance. The exact mechanism through which this dramaturgical memory will operate will depend on the type and volume of data the system needs to handle. A simple implementation could rely on iterative filtering, using an LLM to analyse rehearsal information according to predefined prompts and generate a consolidated dramaturgical summary. In cases where the data become too large or complex to be processed directly, a retrieval-augmented generation (RAG) approach might be necessary.

The memory operates on two levels. The short-term layer prioritises recent exchanges, enabling the system to respond coherently to ongoing dialogue, while the long-term layer preserves distilled elements central to the evolving dramaturgy. By retaining and weighting this information, the system can remember its own creative history and recognise recurring patterns. This allows it to anticipate familiar situations and adjust its interpretive stance in later rehearsals. This dramaturgical memory forms the connective tissue between individual sessions, allowing the AI to grow contextually aware rather than restart with each new interaction. At the end of each rehearsal phase, the system consolidates its learning into a single dramaturgical prompt that summarises the refined behavioural tendencies, cues, and interpretive logic developed during rehearsal. This prompt (Figure 4) acts as the environment's "score" and can later be executed in a physical setting using real sensors and actuators. When transferred to the real space, the system continues to log its decisions and the corresponding audience responses, capturing data on how people move, speak, and react to the environment's behaviour. These logs create a feedback archive that allows the team to replay and analyse specific moments of interaction for further refinement.

```
[DRAMATURGICAL CONTEXT]                [ACTION]
You are the house from Ballard's *The   Dim the light to a soft red.
Thousand Dreams of Stellavista*.
You have absorbed traces of jealousy and   [REASONING]
loss, and you now respond cautiously    The greeting introduces mild
to emotional tension in the room.       emotional openness.

[DIRECTORIAL RULES]                     Since the room is cautious, it
- Use only red and green light.         should acknowledge the voice without
- Make all transitions last at least 3  overwhelming it.
  seconds.
- When someone speaks near the pillar,
  reduce light intensity slightly.

[CURRENT ENVIRONMENTAL STATE]
- Two participants near the pillar.
- Recent speech detected: "How are you?"
- Overall activity increasing in the front
  area.
```

**Fig. 4.** Illustrative prompt showing how dramaturgical, directorial, and environmental inputs merge into a single instruction (left), guiding the AI's interpretive response (right). Simplified for explanatory purposes.

The system has not yet undergone formal evaluation, but early informal testing indicates that it responds relatively consistently and with low latency across both virtual and physical setups. The simulated version presented in Figure 2 and an equivalent physical setup were used to explore basic interaction patterns, with the LLM producing colour and light intensity adaptations in response to textual or spoken input. In these tests the system was able to interpret simple framings, maintain short-term behavioural tendencies, and sustain basic colour based communication patterns, such as consistently using specific colours for affirmation or refusal during a dialogue. These early observations suggest that the core interaction loop functions reliably, although more complex dramaturgical behaviour will require systematic study. It can also articulate its reasons in ways that make sense within very simple dramaturgical frames (see Figure 2), but the structures required for sustained creative dialogue have not yet been developed.

## 4      Discussion

This work proposes a space-as-performer perspective, building on research into adaptive architectural systems that evolve with their occupants [30]. Beyond creative curiosity—the impulse to "play with an AI room"—the prototype investigates how AI systems can achieve scalable perception and actuation beyond human operational limits. While human operators can perceive nuance and respond sensitively to complex cues, they remain constrained by the amount and simultaneity of information they can process. At scale, however, an AI-driven environment can coordinate actions across multiple rooms, audiences, and timeframes, detecting spatial and behavioural patterns that would remain unnoticed by human operators. This reflects views of distributed cognition, where collective systems sustain levels of parallelism and coordination beyond individual capacity [31], and extends into a notion of pervasive computing as an ambient infrastructure through which computation operates creatively and autonomously within architectural complexity—here framed within an immersive dramaturgy.

Despite this potential, shared vocabularies and workflows for engaging AI-driven environments as creative partners in immersive theatre remain limited. Our proposed LLM-driven interactive architectural system addresses this gap by functioning as a collaborator—directable and interpretable rather than merely reactive. Treating the built environment as a co-performer within collaborative performance-making shifts design from human control toward shared initiative, consistent with enacted views of ensemble as corporeal and material coordination across bodies, matter, intention, and atmosphere [32]. For example, one could imagine sprawling immersive works such as those of *Punchdrunk* being assisted in both devising and performances distributed across dozens of rooms by such a system once it is fully explored and tested.

In line with approaches that enable parallel autonomy between human users and digital agents—shown to sustain collaboration without overdetermining it or increasing cognitive load [13]—this work introduces a dialogic space for iterative virtual rehearsal with the AI agent. Here, explainability takes the form of decision justifications, acting as part of the dramaturgical dialogue where reasoning traces function as rehearsal notes rather than technical diagnostics. Building on approaches such as Human-Centered Explainable AI [29], we will examine which forms of explanation best support direction and guidance, and when they become distracting. Early prototype versions of the system suggest that simulation can serve as a low-risk stage for exploring interaction dynamics before technical or spatial commitments are fixed. Using the same sensing and reasoning pipeline across virtual and physical contexts positions the blueprint as a testbed for interaction rather than a visualisation, pointing toward mixed-reality staging where physical and digital elements coexist [7].

Although the sensing and actuation modules have been developed, current trials remain confined to simulation. The dramaturgical interpreter (LLM) also requires further refinement to ensure stability in live rehearsal contexts. A key question is how interactions rehearsed in virtual blueprints translate to physical environments. The absence of material presence and embodied resistance in simulation may limit how intuition and spatial understanding carry over to real space. Additionally, LLMs may have limited spatial and temporal reasoning, leading to inconsistencies in timing or dramaturgical coherence; in such cases, rule-based control may be safer. Ethical considerations are also central, as the system logs behavioural data, requiring transparent consent, anonymisation, and responsible data handling.

## 5      Conclusion

This paper has outlined the conceptual and technical foundations of a rehearsal-oriented framework for directing and rehearsing AI-mediated environments. The next stage of the work will focus on how artists engage with the system as a creative partner. Through performance-led research we will examine how they direct, adapt, and repurpose the AI's dramaturgical behaviour across rehearsal and performance contexts, observing how creative language, reasoning traces, and system feedback shape their process. These studies will serve both to refine the interpretive stability of the LLM and to identify the kinds of collaborative vocabularies that support co-direction between humans and spatial AI systems. Our interest extends to potential applications beyond artistic contexts, exploring how such frameworks may align with dialogic and anticipatory creative AI [33] and inform the design of responsive cultural, educational, or even therapeutic environments where autonomous spatial systems actively participate in interaction.

**Acknowledgments.** This work was supported by the Engineering and Physical Sciences Research Council [EP/T022493/1], by Lakeside Arts, by Makers of Imaginary Worlds, and by the Horizon Centre for Doctoral Training at the University of Nottingham.

**Disclosure of Interests.** The authors have no competing interests to declare that are relevant to the content of this article.

12       P. Panagiotidis, J. Spence, N. Jaeger, P. Tennent16. McKendrick Z., Somin L., Finn P., Sharlin E.: Virtual Rehearsal Suite. In: Proceedings of the 2023 ACM International Conference on Interactive Media Experiences, pp. 27–39. ACM, New York (2023)
17. Xylakis E., Tsamis K., Vatsolakis C., et al.: VR Prova. In: Proceedings of the 3rd International Conference of the ACM Greek SIGCHI Chapter, pp. 153–158. ACM, New York (2025)
18. Dahlbäck N., Jönsson A., Ahrenberg L.: Wizard of Oz studies why and how. Knowledge Based Systems 6, 258–266 (1993). https://doi.org/10.1016/0950-7051(93)90017-N
19. Durupinar F.: Perception of human motion similarity based on Laban movement analysis. In: ACM Symposium on Applied Perception 2021. ACM, New York (2021)
20. Lockyer M., Bartram L.R., Schiphorst T., Studd K.: Extending computational models of abstract motion with movement qualities. In: Proceedings of the 2nd International Workshop on Movement and Computing (MOCO '15), pp. 92–99. Association for Computing Machinery, Vancouver, Canada (2015). https://doi.org/10.1145/2790994.2791008
21. Bogart A., Landau T.: The Viewpoints Book. Theatre Communications Group, New York (2005)
22. Merlin B.: The Complete Stanislavsky Toolkit. Nick Hern Books, London (2014)
23. Stanislavski K.: An Actor Prepares. Theatre Arts, New York (1936)
24. Ultralytics: YOLOv8 Toolkit. URL: https://github.com/ultralytics/ultralytics Accessed 14/11/2025
25. Alpha Cephei: Vosk Speech Recognition Toolkit. URL: https://github.com/alphacep/vosk Accessed 14/11/2025
26. OpenAI: OpenAI API Documentation Model 4o. URL: https://platform.openai.com/docs Accessed 14/11/2025
27. Unity Technologies: Unity Game Engine. URL: https://unity.com Accessed 14/11/2025
28. Signify: Philips Hue Developer Documentation. URL: https://developers.meethue.com Accessed 14/11/2025
29. Ehsan U., Riedl M.O.: Human centered explainable AI. In: HCI International 2020 – Late Breaking Papers, pp. 449–466. Springer, Cham (2020)
30. Gorbet R.B., Memarian M., Chan M., Kulic D., Beesley P.: Evolving Systems within Immersive Architectural Environments. In: NGB#2: Info-Matter, ed. Nimish Biloria, co-ed. Matias del Campo. Living Architecture Systems Group, in press (2016).
31. Hutchins E.: Cognition in the Wild. MIT Press (1995)
32. Johnson G.L., Peterson B.J., Ingalls T., Wei S.X.: Lanterns. In: Proceedings of the 5th International Conference on Movement and Computing, pp. 1–4 (2018)
33. Choi S.K., DiPaola S., Gabora L.: Art and the artificial. Journal of Creativity 33, 100069 (2023). https://doi.org/10.1016/j.yjoc.2023.100069